# Laser-written reconfigurable energy landscapes and programmable Moiré spin textures

*Matteo Panzeri, Piero Florio, Davide Girardi, Joseba Urrestarazu, Giacomo Sala,
Nicola Pellizzi, Matteo Vitali, Marco Madami, Luca Ciaccarini Mavilla, Silvia Tacchi,
Elisa Riedo, Andrea Meo, Vito Puliafito, Mario Carpentieri, Riccardo Tomasello, Efe Ersoy,
Kai Wagner, Patrick Maletinsky, Olivier Boulle, Edoardo Albisetti\*, Daniela Petti\**

M. Panzeri, P. Florio, D. Girardi, G. Sala, N. Pellizzi, M. Vitali, E. Albisetti, D. Petti
Dipartimento di Fisica, Politecnico di Milano, Piazza Leonardo da Vinci 32, Milano, Italy

J. Urrestarazu, O. Boulle
Université Grenoble Alpes, CNRS, CEA, SPINTEC, 38054 Grenoble, France

M. Madami, L. C. Mavilla
Dipartimento di Fisica e Geologia, Università di Perugia, Via A. Pascoli, Perugia, Italy

S. Tacchi
Istituto Officina dei Materiali - Consiglio Nazionale delle Ricerche (CNR-IOM), Sede Secondaria di Perugia c/o Dipartimento di Fisica e Geologia, Università di Perugia, Perugia, Italy

E. Riedo
Department of Chemical and Biomolecular Engineering, Tandon School of Engineering, New York University, Brooklyn, NY, 11201, USA

A. Meo, V. Puliafito, M. Carpentieri, R. Tomasello
Department of Electrical and Information Engineering, Politecnico of Bari, Bari 70125, Italy

E. Ersoy, K. Wagner, P. Maletinsky
Department of Physics, University of Basel, Basel, Switzerland

**Abstract text.** Magnetic textures are central to emerging spintronic and unconventional computing technologies due to their rich dynamics, topological properties and nanoscale dimensions. A major challenge remains achieving tunable, reversible, and spatially resolved control over these textures and their evolution as a function of external stimuli, by spatially reprogramming the magnetic energy landscape that governs their nucleation and stability. Here, we exploit a focused laser-assisted local field cooling technique that establishes a fast, non-contact and scalable platform for grayscale spin texture engineering. By non-destructively controlling the exchange-bias anisotropy with nanoscale resolution in thin-film heterostructures, this approach enables grayscale, reprogrammable control of the local energy profile, which we use to create magnetic patterns with highly controlled hysteresis, field-dependent readability and tunable switching thresholds. Leveraging this capability, we demonstrate information encoding with magnetic field-gated readability, and artificial spin metamaterials, stabilizing spin lattices with field-reconfigurable symmetries and creating artificial Moiré spin textures via the geometric superposition of twisted magnetic potentials. These results establish a versatile, reprogrammable platform that bridges the gap between application-oriented magnetic memory and fundamental studies of emergent order in artificial lattices.

Giacomo Sala's current address is: Department of Quantum Matter Physics, University of Geneva, Geneva, Switzerland.

# 1. Introduction

In magnetic materials, the overall energetic landscape is determined by the combination of several terms arising from the material properties and structure, its geometry and dimensionality, and interfacial phenomena arising from thin-film stacking. This landscape sets the equilibrium magnetization and controls dynamical responses to external stimuli such as magnetic fields or currents. The ability to tailor this landscape at will opens pathways for encoding, storing, and processing information in the magnetic state, forming the foundation of spintronic and magnonic technologies.[1,2] The minima of this energy landscape stabilize the magnetic textures (domains, vortices, bubbles), which act as reconfigurable circuit elements, enabling memory bits,[3,4] logic gates,[4,5] and hardware primitives for neuromorphic[6,7] or probabilistic computing[8,9] owing to their intrinsic stability and nanoscale size. In ultrathin heterostructures with perpendicular magnetic anisotropy (PMA), interfacial effects become particularly dominant. The broken inversion symmetry at ferromagnet/heavy-metal interfaces gives rise to a sizable Dzyaloshinskii–Moriya interaction (DMI),[10–14] which can stabilize nanoscale chiral textures such as Néel domain walls and skyrmions.[15,16] Exchange bias coupling with an adjacent antiferromagnetic layer further enriches this landscape, introducing a unidirectional anisotropy term that acts as a built-in effective field.[16–18] Through the interplay of DMI, anisotropy, and exchange bias, ultrathin PMA heterostructures can host a wide range of reconfigurable spin textures. For instance, magnetic skyrmions are chiral, topologically nontrivial spin textures with nanoscale diameters whose energetic stability is set by the local balance of exchange, anisotropy, dipolar and interfacial DMI contributions.[19–21] Experimentally, skyrmions may appear either as isolated, stochastic entities or as ordered,[17,22–24] close-packed hexagonal crystals.[25–27]

To transition from fundamental studies to practical applications, it is essential to achieve spatially resolved, quantitative, and reversible programming of the underlying energy potential that stabilizes magnetic textures. Various approaches have been explored to achieve local modification of the magnetic landscape, including electron beam lithography,[28] laser writing,[29–33] additive fabrication of nanostructures,[34,35] ultrafast optical or X-ray excitation,[36–38] local ion or magnetic-tip-induced changes.[39–41] While these techniques provide spatial selectivity, they generally do not allow a quantitative, reversible, grayscale tuning of the magnetic energy landscape, which is key for exploring new physical phenomena and next-generation computing. Typically, they rely on permanent structural changes such as local crystallization, oxidation, or interdiffusion. This fundamentally limits the rewritability of the device and prevents the dynamic reconfiguration of the lattice symmetries on the same active area. A different solution, enabling reconfigurable patterning, is represented by thermally assisted magnetic scanning probe lithography (tam-SPL),[42–45] which allows sub-10 nm resolution. However, its inherently serial, contact-based nature presents a fundamental bottleneck for speed, large-area scalability, and integration. Moreover, achieving precise grayscale control of the energy density over macroscopic areas remains challenging with scanning probes due to tip wear and thermal drift.

Therefore, developing a straightforward and fully reversible method to locally shape magnetic energy landscapes in perpendicular magnetic multilayers in a quantitative way, without altering structural integrity, remains a key challenge.

In this work, we introduce a laser-assisted local field cooling technique that enables grayscale, reprogrammable control over key magnetic parameters in exchange-biased perpendicular magnetic heterostructures hosting skyrmions.[30,42,44,46] This approach allows for quantitative, grayscale, and spatially resolved programming of the unidirectional anisotropy, effectively defining magnetic potentials that govern texture nucleation, stability, geometry, and magnetization dynamics. By tailoring these potentials, we demonstrate the nucleation of complex domains with spatially tailored hysteresis for information encoding, and the reversible transformation between distinct spin-texture lattice geometries such as hexagonal, square, and kagome. Crucially, the non-destructive nature of our field-cooling protocol allows for the additive superposition of distinct magnetic potentials. We exploit this unique feature to generate artificial Moiré spin textures by sequentially writing twisted or mismatched lattices.

## 2. Results
### 2.1. Programmable Grayscale Anisotropy Landscape
*2.1.1. On-Demand Nanoscale Exchange Bias Profile*

We implemented the laser-assisted local field cooling technique on an exchange-biased Ta 5/IrMn 6.5 (Ru 2)/Ta 0.2/CoFeB 1/MgO 2/Ta 2 (nm) heterostructure grown by magnetron sputtering to demonstrate the grayscale programming of exchange bias (EB) and exchange-bias anisotropy landscapes with sub-µm spatial resolution. The process, illustrated in **Figure 1a**, involves locally heating the material close to its blocking temperature[42,44,46] with a focused 405 nm continuous wave laser beam, in the presence of a static external magnetic field. This temporarily weakens or destroys the local exchange bias, which, upon cooling, is reset along the applied field direction.

**Figure S1** shows the sample's multilayer structure (Figure S1a) as well as the representative hysteresis loop (Figure S1b), from which we determine the global exchange bias value of 11 mT, and the outwards direction of the magnetization at remanence. We then irradiated the sample in ambient conditions, in the presence of an inward external static field of around 60 mT generated by permanent magnets, to create a complex spatially varying energy landscape. This was achieved by exposing a double-conical gradient pattern, where the laser power was modulated continuously from 3.4 to 5.2 mW (**Figure 1c**). The overall pattern, measuring 50x50 µm$^2$, was exposed in a raster-scan fashion with a pixel size of 100 nm and an exposure time of 10 µs per pixel. Magneto-optical Kerr effect (MOKE) microscopy images acquired during both negative-to-positive ("rise", **Figure 1f**) and positive-to-negative ("fall", **Figure 1g**) sweeps of the external field were used to probe the field-dependent switching of the magnetization in both irradiated and non-irradiated regions. From the Kerr data, spatial maps of the switching fields were extracted for both field sweep directions (**Figures 1h and 1i**), highlighting how the spatial modulation of laser power translates into a spatially varying shift of the magnetization switching fields. Further analysis yielded the spatial distributions of the exchange bias field ranging from -11 mT to +11 mT (Figure 1d) and the corresponding exchange-bias anisotropy energy density across the pattern ranging from -6 to +6 kJ/m$^3$ (**Figure 1e**), confirming the presence of a spatially varying $H_{EB}$ leading to a smooth magnetic energy landscape. The direct impact of the continuous modulation of the laser patterning power is evident by considering the local hysteresis loops (**Figure S1c**) and the local exchange-bias anisotropy $K_{EB}$ (**Figure 1b**) values. Both plots confirm the smooth, continuous transition as a function of the laser power, from the pristine (unpatterned) film properties with a loop shifted towards negative field values, to a complete shift of the hysteresis loop towards positive field values and a complete reversal of the exchange-bias anisotropy direction. Importantly, the irradiation selectively modifies the unidirectional anisotropy, while leaving the other magnetic properties unchanged, allowing to quantitatively design the magnetic energetic landscape. This stems from the fact that the blocking temperature of the system, and therefore the temperature reached upon laser heating, is well below the temperature leading to structural modifications such as interdiffusion at the interfaces. The plot of the exchange-bias anisotropy $K_{EB}$ as a function of the laser power of Figure 1b reveals an initial power threshold of 3.7 mW, followed by an approximately linear variation of the exchange-bias anisotropy constant until saturation is reached at 4.7 mW, where a complete reversal is achieved. By re-mapping this $K_{EB}(P)$ calibration onto the design, we are able to quantitatively re-set the magnetic anisotropy value point-by-point, therefore controlling not only the spin texture at remanence, but also its whole spatial evolution as a function of the external field. This continuous variation of the exchange-bias anisotropy with the power is well described by a thermally activated process involving the progressive switching of antiferromagnetic grains of different sizes as the temperature increases.[47] More generally, beyond grayscale gradients, the same raster-writing protocol enables single-pass fabrication of complex magnetic patterns with arbitrary geometry (see Figure S1d).

*2.1.2. Field-Reconfigurable Functional Magnetic Patterns and Encoding.*

To further highlight the versatility of the technique, we applied it beyond continuous anisotropy landscapes, demonstrating magnetic information encoding and the realization of field-reconfigurable magnetic structures. As a first example, we patterned a "gradient tree" structure (**Figure 2a**), where the spatially varying laser power produces a corresponding modulation in the exchange bias field. The resulting spatial $H_{EB}$ profile (**Figure 2b**) directly reflects the tree layout, and Kerr microscopy measurements reveal distinct remanent states at zero external field, depending on the field sweep direction (**Figures 2c–d**). Importantly, we are able to encode two stable remanent states, whose configuration depends on whether the system is first saturated from negative or from positive fields. This demonstrates a form of bistable magnetic information encoding, where the same structure can host two robust and spatially different non-volatile zero-field states. As a proof-of-concept, we encoded a QR code pattern linking to our laboratory's website, which can be machine-read magneto-optically only in specific ranges of external magnetic field which are determined by the irradiation power map. A single-power layout was designed, as shown in the top of **Figure 2e**. The readability diagram, displayed in **Figure 2f** and obtained from Kerr microscopy, shows that the code functions as a readable QR code only during specific field intervals. For the lower powers (4.5 and 4.7 mW), the readability window is quite narrow, and the QR code is readable at remanence (zero field) only during a negative-to-positive field sweep ("rise only"), and around 15 mT during a positive-to-negative field sweep ("fall only"). In contrast, for higher laser powers, the readability window becomes broader, allowing the QR code to be readable in both remanent states at zero field. We then designed a more complex, "two-color" QR code (Figure 2e, bottom). This approach, by using two different laser powers for patterning, creates regions with distinct exchange-bias anisotropies and therefore different magnetic switching fields, consistent with the results in Figures 1e and 1h-i. The two-color codes exhibit an expanded field-dependent readability, as summarized in the diagram of **Figure 2g**, producing regions with different switching thresholds, allowing the code to be reconstructed over a wider field range. Kerr microscopy images of the 4.5-4.9 mW two-color code at different external fields (**Figure 2h**) illustrate this effect: different parts of the QR code emerge progressively as the external field is swept, reflecting the local magnetic switching thresholds. For instance, the QR code in Figure 2h is fully readable only in the remanent state following the rising (negative-to-positive) field sweep, highlighted by the green frame, while the remanent state after the falling (positive-to-negative) sweep remains unreadable. This demonstrates how the use of grayscale magnetic landscapes allows reversible multi-level spatial encoding of information, which can be retrieved only under specific external field conditions, and is otherwise either unreadable or magnetically invisible at different fields.

## 2.2 Reconfigurable Chiral Magnetic Lattices

*2.2.1. Chiral Magnetic Lattices with Deterministic Geometry.*

We next demonstrate how our laser-assisted local field cooling method can be used to directly write magnetic energy landscapes that can stabilize chiral domains. For this purpose, we employed a multilayer of Ta 5/Pt 3/Co 0.3/NiFe 1.9/IrMn 4.2/Pt (nm) (**Figure S2a**). The presence of an exchange bias field of 55 mT (**Figure S2b**) and the interfacial Dzyaloshinskii-Moriya interaction in this system enables the formation of chiral spin textures such as magnetic skyrmions.[22]

By locally tailoring the exchange bias field with laser patterning, we imprint a spatially programmable magnetic potential that dictates the nucleation sites and stability of these spin textures. Building on this principle, we realized ordered magnetic lattices with arbitrary geometries. As shown in **Figures 3a-c**, we successfully patterned magnetic lattices with simple square, hexagonal, and kagome geometry, by using a laser power of 7.4 mW. The square lattice has a lattice parameter of 1 μm, while the hexagonal and kagome lattices have a lattice parameter of 1.5 μm. For all these geometries, the chiral (bubble) domain dimensions are approximately 550 nm. The insets in these figures display the corresponding 2D FFTs, which confirm the periodicity and symmetry of the programmed lattices. Importantly, this approach enables the independent control of lattice geometry and lattice parameter. **Figure S3** further demonstrates this tunability across geometries: for a given lattice geometry, the lattice parameter can be changed from 2.5 μm, where the spin textures are separate and non-interacting, down to values where the separation is comparable to the feature size (about 500 nm), leading to partial spatial superposition.

We further extended this approach to create field-reconfigurable magnetic lattices, where distinct sublattices with different switching thresholds are encoded by varying the laser writing power. For instance, a centered square magnetic lattice was fabricated using a tailored power map (**Figure 3d**), where each lattice dot corresponds to a single-shot laser pulse with an exposure time of 10-20 μs. MFM images reveal a clear field-driven transformation of the spin textures. At remanence, a centered square geometry is visible, with feature sizes of 1.2 μm and 800 nm for the high-power and low-power sublattices, respectively (**Figure 3e**). When an external field of -4 mT is applied, the features patterned at the lower powers disappear, transforming the structure into a simple square lattice (**Figure 3f**). This reconfiguration arises from the engineered differences in local magnetic potential, which set distinct switching thresholds for each sublattice. The shrinking of the features is attributed to the Gaussian profile of the laser spot, where the tails of the pulse have a lower effective power, leading to a local magnetization reversal at lower applied fields, consistent with our earlier results. Similarly, by applying the same laser power map to a hexagonal geometry (**Figure 3g**) the hexagonal lattice observed at remanence (**Figure 3h**) transitions to a kagome lattice at the same applied field of -4 mT (**Figure 3i**).

Finally, we show that the energy landscapes are reversible and rewritable. By repeating the exposure under opposite field directions, previously written lattices can be selectively erased and replaced with new ones. **Figures S4a-c** shows the different stages of the writing, erasing and re-writing process. A square lattice was first written, then partially erased by exposure with an opposite external magnetic field, and finally overwritten with a hexagonal lattice in the same region. This three-step sequence highlights the non-destructive, fully reprogrammable nature of our method, distinguishing it from permanent techniques such as ion-beam or X-ray irradiation.

Together, these results demonstrate that laser-assisted local field cooling enables precise, grayscale and reprogrammable spatial control over exchange-bias landscape in magnetic thin films. This capability allows for deterministic design and field-driven reconfiguration of artificial chiral spin textures.

## 2.3 Spin Wave Dynamics in Artificial Lattices

The dynamical properties of the magnetic lattices prepared in the Ta/Pt/Co/NiFe/IrMn/Pt multilayer were investigated by means of Brillouin Light Scattering (BLS) measurements, using a ~40 μm diameter laser spot that averages the response over several dozen lattice sites. In particular, we studied a hexagonal lattice written with a laser power of 7.6 mW and having a period of 1 μm. BLS measurements were carried out in Damon-Eshbach configuration as a function of the in-plane applied magnetic field, $H_{in}$, from zero up to 300 mT, depicted in the inset of **Figure 4a**. Representative BLS spectra acquired at different values of the applied field are shown in Figure 4a. As seen on both the Stokes and the Anti-Stokes sides of the spectra, a single peak is observed, whose frequency increases with increasing magnetic field. Note that for the unpatterned multilayer no BLS signal is detected at remanence, since the BLS cross-section vanishes when the film magnetization rotates out-of-plane. The average value of the Stokes and Anti-Stokes frequencies as a function of $H_{in}$ is reported in **Figure 4c** (red squares). Starting from zero field, the mode frequency remains approximately constant at about 1.8 GHz up to around 20-25 mT; for higher values, it increases linearly with $H_{in}$. Similar measurements were performed by varying the in-plane direction of $H_{in}$ with respect to the hexagonal lattice. We found no measurable change in the mode frequency, indicating a negligible interaction between neighboring textures (data not shown). The observed resonance is therefore consistent with the single-texture breathing mode, as further supported by the micromagnetic simulations discussed below.

To gain a deeper understanding of the experimental results, we performed micromagnetic simulations (see the "Methods" section for details of the model). The strength of the i-DMI was set to $D = 0.35$ mJ/m$^2$ estimated from BLS measurements on the unpatterned multilayer (see **Figure S6**). We then carried out a systematic study by varying the perpendicular magnetic anisotropy constant $K_u$ in the range 290 kJ/m$^3$ < $K_u$ < 350 kJ/m$^3$ (see **Figure S7**). We set $K_u$ = 320 kJ/m$^3$, i.e. the minimum anisotropy such that the system is perpendicularly magnetized in zero field, in agreement with the hysteresis loop reported in Supplementary Figure S2. The relaxed magnetic configuration at remanence displays a chiral Néel-type domain wall (see **Figure 4b**).

Figure 4c compares the BLS data with the simulations, showing a good qualitative and quantitative agreement: an almost constant frequency at low $H_{in}$, followed by a linear increase for larger fields. To understand the reason behind the change in the frequency behavior, we analyze both the static micromagnetic textures at different fields, shown in **Figure 4d**, and the spatially resolved excitation modes.

At zero field, the equilibrium configuration corresponds to a standard Néel-type skyrmion, and the excited dynamics are associated with a skyrmion breathing mode (see Figure S7d). However, confinement induced by the exchange bias leads to two differences with respect to an unconfined skyrmion[37,48,48–53]: (*i*) a higher frequency, and (*ii*) a non-uniform power distribution (see Figure S7d).

For increasing fields up to $H_{in}$ = 50 mT, we observe a progressive tilt of the magnetization along the +x direction (see snapshots at 10, 30 and 50 mT), while the excited breathing-mode frequency remains nearly constant. This behavior is ascribed to the magnitude of the field, which is not sufficiently large to overcome the skyrmion topological protection.

For $H_{in}$ > 50 mT, the field induces a larger tilt of the magnetization, approaching in-plane saturation. At the same time, the skyrmion transforms into a bubble, losing its topological protection. In this regime, the excited dynamics are no longer associated with a breathing mode, but with a ferromagnetic mode whose field dependence follows the expected Kittel-like behavior.[54] We attribute the small difference (~0.2 GHz) between the experimental and simulated frequencies for $H_{in}$ > 50 mT to the fact that the simulation models a finite-size 750 nm x 750 nm film rather than an extended film, leading to an underestimation of the magnetostatic contribution. Similar frequency values have also been obtained for rectangular and hexagonal skyrmion lattices (not shown), indicating that the resonant response corresponds to the breathing mode of an individual spin texture and that the interaction between neighboring elements is negligible.

## 2.4 Moiré Spin-textures
### 2.4.1. Twisted Moiré Lattices

To extend the concept of programmable magnetic potentials, we investigated how energy landscapes can be superimposed to generate emergent long-range ordered spin textures. In this framework, Moiré magnetic lattices arise not simply from a direct superposition of pre-existing spin-texture patterns; rather the superimposed magnetic energy landscapes encoded by laser writing define a composite pinning potential in which the spin textures subsequently nucleate and stabilize, yielding the emergent Moiré ordering. Two main strategies were implemented to realize this concept: overlapping identical lattices with a relative twist, and superimposing lattices with different periodicities. **Figure 5a** and **Figure 5b** illustrate these approaches schematically. We implemented these strategies through a two-step laser-assisted patterning process: in the twisted case, the same lattice geometry was written twice with a controlled angular misalignment achieved by rotating the laser scan path. As the applied field during the second exposure in overlapping areas is in the same direction of the exchange bias written in the first pattern, the two landscapes coexist and add constructively, enabling deterministic formation of superlattices through their interference. **Figures 5c–e** show MFM images of Moiré spin texture formed by rotational misalignment, patterned with a laser power of 6.8 mW and with tilt angles of 5.5° (Figure 5c), 7° (Figure 5d), and 8.5° (Figure 5e). The periodicity P of this new lattice is dependent on the angle of rotation $\vartheta$ between the two initial lattices according to the relation $P = p/2\sin(\vartheta/2)$, where $p$ is the periodicity of the original lattice, in this case 2 μm. The formula yields expected Moiré periodicities of 20.8 μm, 16.4 μm and 13.5 μm respectively. These values are in good agreement with our experimentally measured periodicities of 20.9 μm, 16.3 μm and 13.7 μm. Moreover, the corresponding 2D Fast Fourier Transforms (insets) not only reproduce the reciprocal lattice of the original arrays, but also reveal additional satellite peaks surrounding each primary spot. These satellites correspond to the long-range Moiré modulation, appearing at much smaller reciprocal vectors, and thus directly confirm the coexistence of the original lattice periodicity with the emergent Moiré pattern.

*2.4.2 Mismatched Moiré Lattices*
The flexibility in designing custom lattices allowed us to explore other means of generating long-range Moiré patterns. In a complementary set of experiments, we explored the effect of mismatched periodicities between the two lattices. In this case, two distinct lattice layouts with mismatched periodicities were sequentially patterned. The MFM images in **Figures 5f–h** show the resulting patterns for various combinations of periodicities. Specifically, we generated Moiré patterns by superimposing lattices patterned with a laser power of 6.8 mW, with periodicities of 2.5 μm and 1.5 μm (Figure 5f), 2.5 μm and 3.5 μm (Figure 5g), and 2.5 μm and 3 μm (Figure 5h). The resulting Moiré periodicity P is determined by the mismatch $\delta p$ in the initial lattice parameter $p$ following the formula $P = {p^2}/{\delta p} + p$. The dashed lines in these figures highlight the Moiré periodicities of 3.1 μm, 8.5 μm, and 14.6 μm, which are in reasonable agreement with the predicted values of 3.8 μm, 8.8 μm and 15 μm. While the discrepancies are slightly larger than in the twisted case, we attribute these primarily to measurement errors due to the smaller long-range periodicity. The accompanying 2D FFTs (insets) confirm the emergence of new periodic components not present in the individual lattices, consistent with Moiré-type interference. Moreover, NV magnetometry measurements[55,56] on similar patterns provide direct nanoscale evidence of the internal structure of the Moiré lattice sites (**Figure S5**). Specifically, the data shows that low laser powers stabilize clusters of small skyrmions whose density increases with the laser power, consistent with a gradual shift of the hysteresis loop. On the other hand, above a certain power threshold these clusters merge into single, larger chiral domains. This confirms that the high-power writing used for the lattices in Figures 3 and 5 creates single chiral domains, which also justifies the use of a single-domain model in our micromagnetic simulations (Figure 4).
Together, these results establish Moiré spin textures as emergent superstructures formed by the controlled superposition of programmable magnetic energy landscapes. This approach goes beyond structural templating or lithographic methods, providing a reconfigurable route to engineer complex magnetic superlattices and emergent spin textures.

## 3. Conclusions

In this work, we have introduced a laser-assisted local field cooling technique that establishes a fast, non-contact, and scalable platform for the deterministic engineering of magnetic energy landscapes in exchange-biased thin-film heterostructures. By leveraging the control of the exchange-bias anisotropy with a focused optical source, we achieved quantitative, spatially resolved, and grayscale tailoring of key magnetic parameters, such as the exchange-bias anisotropy energy density. This allows to encode quantitative spatially programmable energy landscapes that govern the stability and dynamics of spin textures.

We exploited the versatility of our technique to demonstrate advancements in two distinct areas. First, for functional device applications, we realized the encoding of functional, field-reconfigurable magnetic security patterns, exploiting the unique ability to engineer magnetic potentials and spatially tailor hysteresis loops and switching thresholds.

Second, in the realm of artificial metamaterials, we demonstrated the patterning of magnetic lattices with reconfigurable symmetries, switching between hexagonal, square and kagome geometries, as well as the tunability of the lattice parameter and of the feature dimensions. The reprogrammability of our approach is exemplified by the selective erasure and rewriting of magnetic configurations.

In addition, we showcase the ability to create complex artificial Moiré spin textures, formed by the interference of twisted or mismatched magnetic potentials. These superlattices emerge as collective states, which can be precisely controlled by the laser parameters.

These results establish laser-assisted magnetic patterning as a powerful and highly versatile platform for deterministic, reversible design of smooth energy landscapes and spin textures. By enabling grayscale, reprogrammable, and high-resolution control over magnetic energy landscapes, the approach provides a reconfigurable route to create complex superlattices and functional patterns. This capability opens up new avenues for applications in spintronics, magnonics, and unconventional computing paradigms.

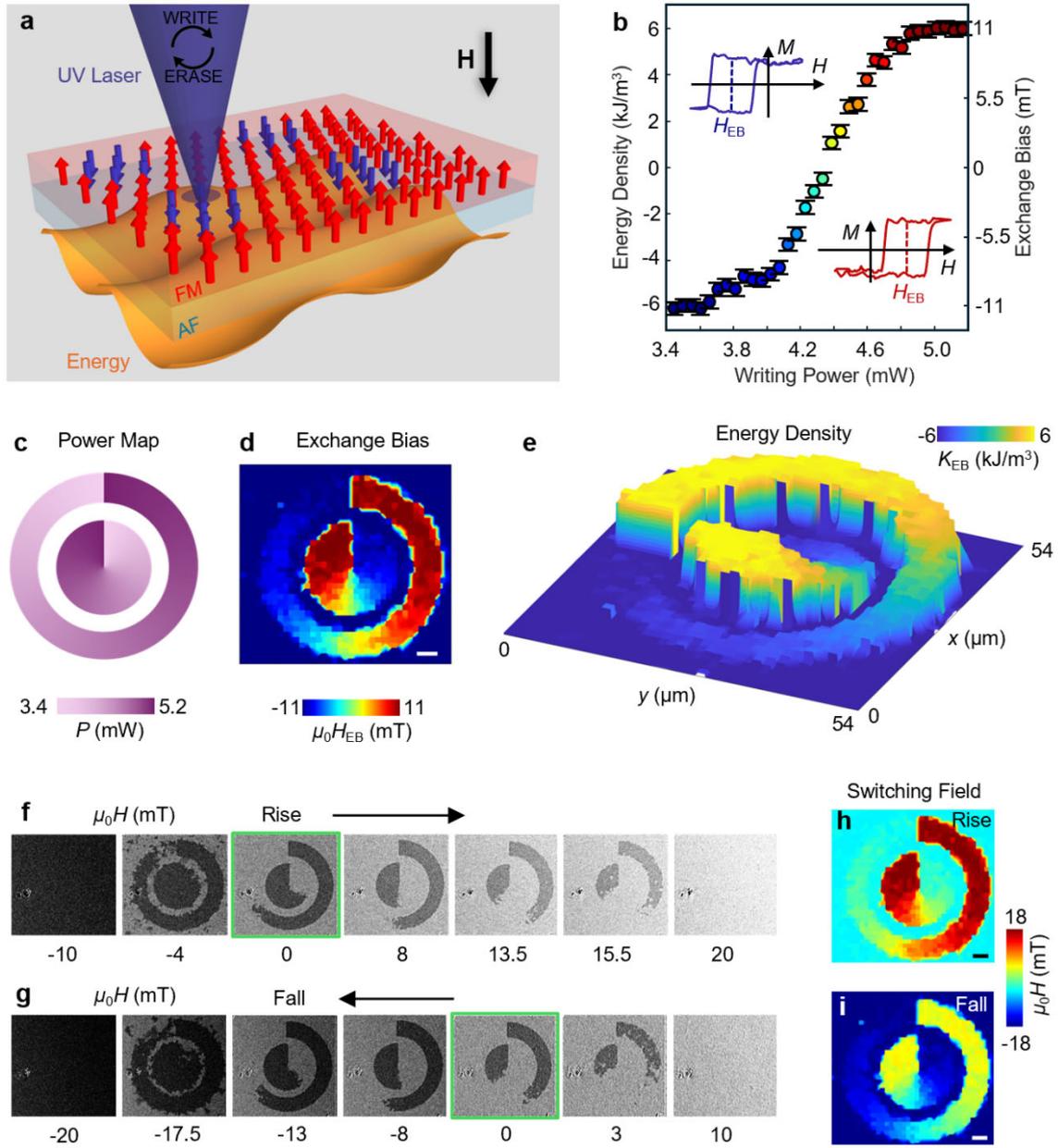

**Figure 1. Quantitative programming of grayscale anisotropy landscapes.** a) Schematic of local field cooling. A focused laser heats the material above the blocking temperature in the presence of an external magnetic field. Upon cooling, the local exchange bias is reset to align with the applied field, enabling programmable energy landscapes. b) Exchange-bias anisotropy energy density (left axis) and corresponding exchange bias field $\mu_0 H_{EB}$ (right axis) as a function of laser power for the double-conical gradient pattern in (c). The error bar is ±0.275 kJ/m³. Representative local hysteresis loops are shown as insets: top-left (negative exchange bias) and bottom-right (positive exchange bias). c) Schematic layout and laser power distribution of the double-conical gradient pattern. d) Spatial map of the exchange bias field $\mu_0 H_{EB}$ obtained from Kerr microscopy (color scale in mT). The scale bar is 5 μm. e) 3D map of the exchange-bias anisotropy energy density $K_{EB}$ extracted from Kerr microscopy. Blue and yellow denote opposite anisotropy directions. f-g) Kerr microscopy images of the patterned structure during a magnetic field sweep: (f) increasing field (rise), (g) decreasing field (fall). h-i) Spatial maps of the magnetization switching fields extracted from Kerr microscopy: (h) increasing sweep (rise), (i) decreasing sweep (fall). Scale bar: 5 μm.

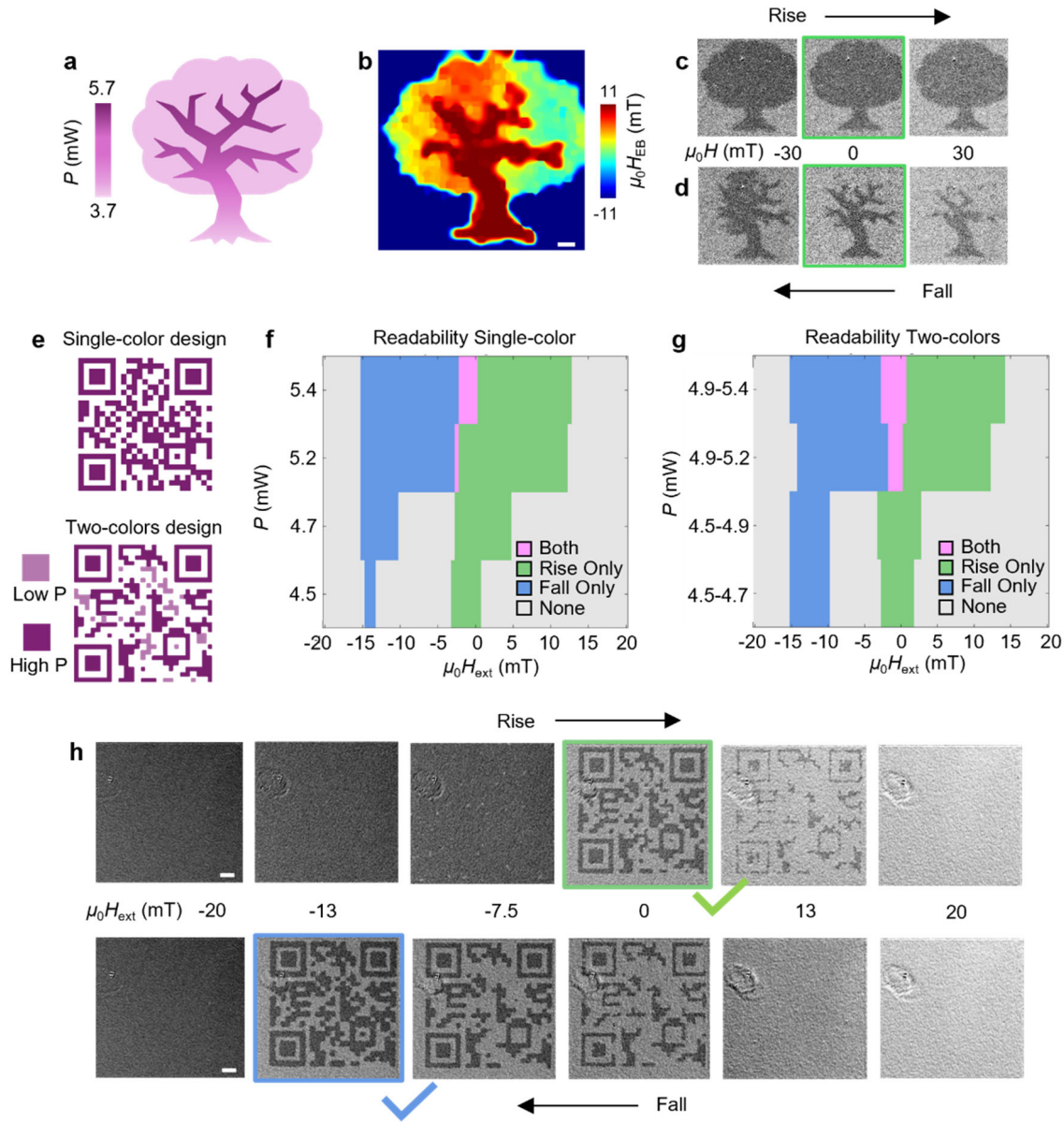

**Figure 2. Magnetic encoding and field-reconfigurability of functional patterns.** a) Schematic layout and laser power distribution for the gradient tree pattern. b) Spatial map of the exchange bias field extracted from Kerr microscopy of the gradient tree. Scale bar: 5 μm. c-d) Kerr microscopy images of the gradient tree during magnetic field sweeps: (c) increasing sweep (rise) and (d) decreasing sweep (fall). The remanent states (zero external field), highlighted by green frames, differ between the two sweeps. These bistable zero-field states are referred to as the "summer tree" (remanence after the increasing sweep) and the "winter tree" (remanence after the decreasing sweep). e) Schematic layouts of magnetically encoded QR codes. Top: single-color (single laser power). Bottom: two-color (two distinct laser powers). Both patterns encode a functional QR code (linking to the PhyND lab website). f-g) Readability diagrams obtained from Kerr microscopy for the single-color (f) and two-color (g) QR codes. "Rise only" indicates the field range where the pattern is fully readable during the field sweep from negative to positive values. "Fall only" indicates the range where the pattern is readable during the field sweep from positive to negative values. h) Kerr microscopy images of the two-color QR code (4.5 and 4.9 mW) at different external fields. The full code is readable only within specific, narrow field ranges. Readable frames are highlighted in green (rise sweep) and blue (fall sweep). Scale bar: 5 μm.

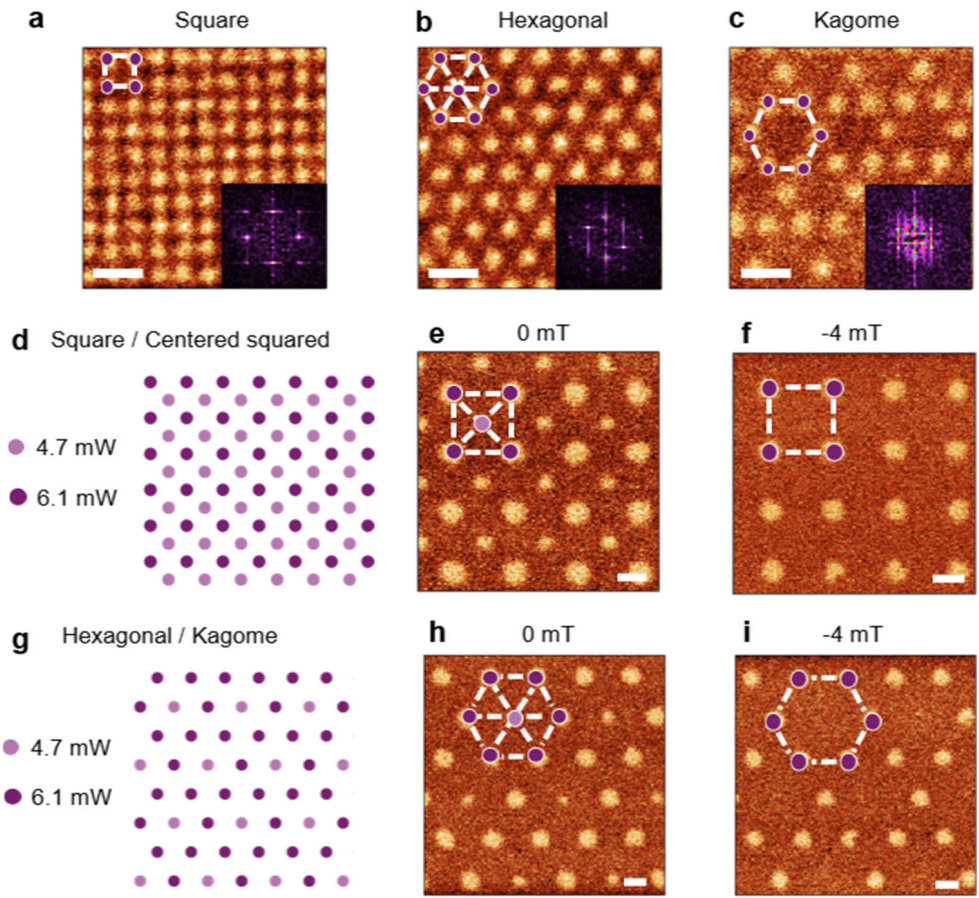

**Figure 3. Programmable magnetic lattices and field-reconfigurable symmetry.** a-c) Magnetic Force Microscopy (MFM) images of magnetic lattices in distinct geometries: (a) simple square, (b) hexagonal, and (c) kagome (scale bars: 2 μm). Insets show the corresponding 2D Fast Fourier Transforms (FFTs), confirming the symmetry of each lattice. d) Schematic layout and laser power distribution for a centered square magnetic lattice. e-f) MFM images of the centered square lattice at remanence (e) and after applying an out-of-plane external field of −4 mT (f), reconfiguring into a simple square lattice. g) Schematic layout and laser power distribution for a hexagonal magnetic lattice patterned using two laser powers. h-i) MFM images of the hexagonal lattice at remanence (h) and after applying an out-of-plane external field of −4 mT (i), showing a reconfiguration into a kagome lattice (scale bars: 3 μm).

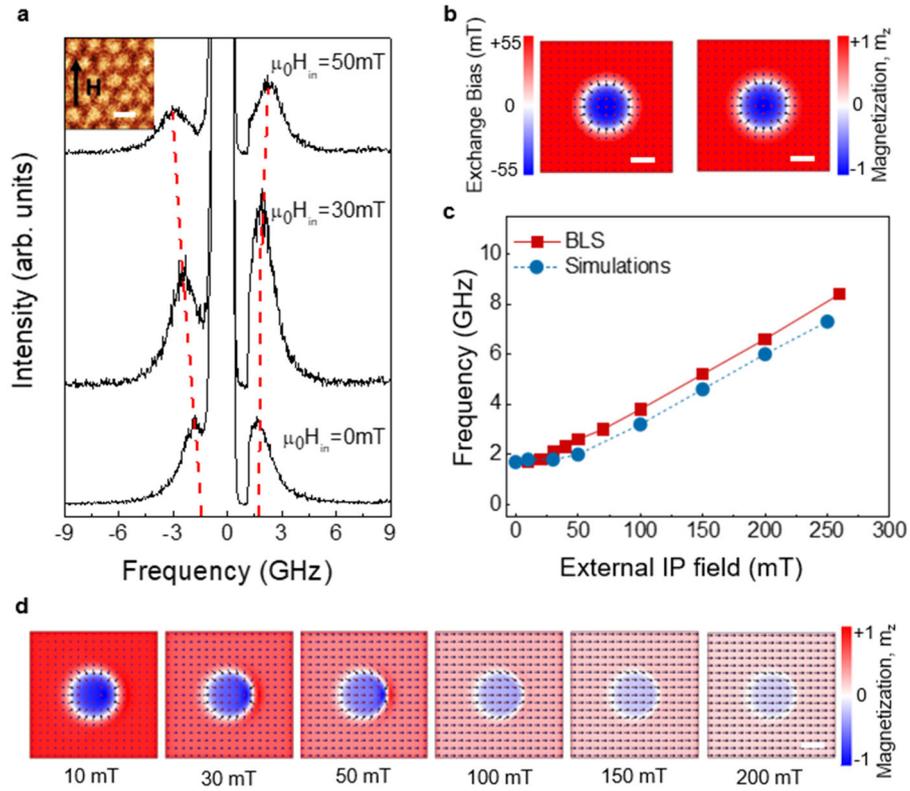

**Figure 4. BLS measurements and micromagnetic simulation results.** a) Wave-vector resolved Brillouin Light Scattering spectra measured at different values of the applied in-plane magnetic field shown in the inset. Inset: Magnetic Force Microscopy image taken at remanence for the hexagonal lattice, having a period of 1 μm, written with a laser power of 7.6 mW on the Ta/Pt/Co/NiFe/IrMn/Pt multilayer stack. Scale bar: 1 μm. b) Top view of spatially varying local exchange bias field (left) and relative relaxed configuration (right) for a single localized spin texture for max($|H_{EB}|$) = 55 mT, as in the experiments. The color bar represents the amplitude of the z-component of the field (left) and magnetization (right). Scale bars: 260 nm. c) Average frequency of the experimental Stokes and Anti-Stokes peaks (red squares) and simulated breathing mode frequency (blue dots) as a function of the applied in-plane magnetic field $H_{in}$. d) Snapshots of the spatial distribution of the z-component of the magnetization for the different fields: $H_{in}$=10, 30, 50, 100, 150, 200 mT. Palette and scale are the same as in b).

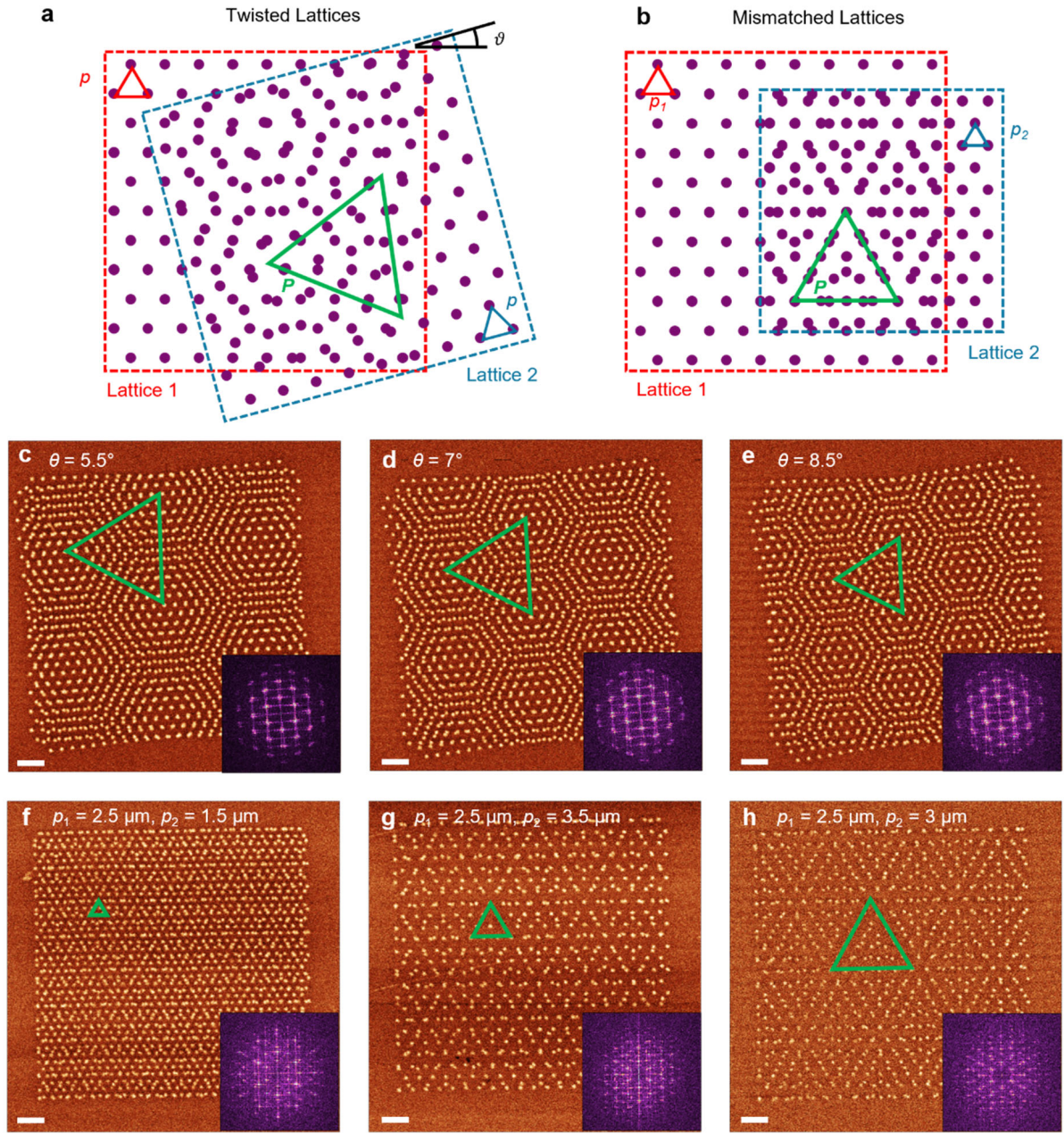

**Figure 5. Generation of Moiré spin textures via geometric superposition of magnetic potentials.** a-b) Schematics of Moiré lattice formation: (a) superimpostion of two identical magnetic lattices with periodicity p and a relative rotation angle θ; (b) superimposition of two magnetic lattices with different periodicities $p_1$ and $p_2$. In both panels, the emergent Moiré periodicity P is marked by the green line. c-e) MFM images of Moiré magnetic lattices generated by rotational misalignment, with relative tilt angles of 5.5° (c), 7° (d), and 8.5° (e). The emergent Moiré periodicity P is marked by the green line (scale bar: 5 μm). Insets: corresponding 2D Fast Fourier Transforms (FFTs), revealing the Moiré pattern in reciprocal space. f-h) MFM images of Moiré lattices generated by superimposing lattices with mismatched periodicities: 2.5 μm and 1.5 μm (f), 2.5 μm and 3.5 μm (g), and 2.5 μm and 3 μm (h). The emergent Moiré periodicity P is marked by the green line (scale bar: 5 μm). Insets: corresponding 2D FFTs, highlighting the emergent Moiré patterns.

## 4. Experimental Section/Methods
*Sample Fabrication:*
The Ta 5/IrMn 6.5(Ru 2)/Ta 0.2/CoFeB 1/MgO 2/Ta 2 (nm) heterostructure was grown via magnetron sputtering, using an AJA ATC Orion 8 system. For conductive materials like Ta, CoFeB, and IrMn, DC mode was employed, while for magnesium oxide (MgO), RF mode was used. The architecture of the Ru-intercalated IrMn layer was realized as a sequential deposition of very thin layers of the two materials, which mixed during growth and the subsequent annealing, forming an alloy with finer crystalline grains. After deposition, the sample underwent a global field cooling. This treatment was carried out in a dedicated system under vacuum, heating the sample to 200°C for 30 minutes in a magnetic field of 4 kOe, applied perpendicularly to the film plane. This step was essential for simultaneously setting two key magnetic properties: the perpendicular magnetic anisotropy (PMA) of the CoFeB layer and the direction of the exchange bias field provided by the doped IrMn layer.

The sample Ta 5/Pt 3/Co 0.3 /NiFe 1.9/IrMn 4.2/Pt (nm) was fabricated using magnetron sputtering deposition on a 100 mm diameter silicon substrate. The selection of optimal thicknesses was made following the characterization of a sample with controlled thickness gradients for the $Ni_{80}Fe_{20}$ and $Ir_{20}Mn_{80}$ layers through "off-axis" deposition, which allows for systematic variation of the thicknesses along the x and y directions of the wafer. After deposition, the sample underwent a thermal annealing treatment at 200°C for 2 minutes under the application of a 570 mT external magnetic field oriented perpendicularly to the film plane to establish exchange bias. The optimization of the thicknesses for the studied sample led to a specific choice: the NiFe thickness was set to 1.9 nm, positioning it near the perpendicular-to-planar magnetization transition, to reduce the dipolar energy and favor small-sized skyrmions. The IrMn thickness was optimized to 4.2 nm to maximize the exchange bias field while simultaneously decreasing the coercivity.[22]

*Direct Laser Writing:* Laser-assisted patterning was performed using a NanoFrazor Explore system (Heidelberg Instruments Nano AG), equipped with a continuous-wave (CW) semiconductor diode laser operating at a wavelength of 405 nm. The laser is focused onto the sample surface through a 20× objective lens, yielding a nominal spot diameter of approximately 1.2 µm. Patterns are converted into arrays of 100 nm × 100 nm pixels and written by raster-scanning the sample with a high-precision piezoelectric stage at a constant velocity determined by the ratio of the pixel size to the pixel time. Each pixel was selectively exposed to the laser beam for a fixed duration, either 10 µs or 20 µs, depending on the sample. Specifically, 10 µs exposures were used for the patterning of the Ta/IrMn/Ru/Ta/CoFeB multilayer (Figure 1 and 2), while 10-20 µs exposures were used for Ta/Pt/Co/NiFe/IrMn/Pt multilayer (Figure 3 and 5). Due to the Gaussian profile of the focused laser beam and the substantial overlap between adjacent spots (as the laser spot size exceeds the pixel size), the local energy deposited at each pixel is a convolution of multiple neighboring exposures. In contrast, for the lattice layouts, each lattice site was defined as a single pixel in the geometry, corresponding to a single-shot exposure.

*Magnetic Force Microscopy (MFM):* MFM measurements were carried out using a Park Systems NX10 instrument operating in lift mode, equipped with PPP-MFMR magnetic probes (Nanosensors). Post-processing of the MFM data involved polynomial background subtraction and correction for line scan artifacts. Measurements under external magnetic fields were performed using a Caylar Magnetic Field Module integrated with the AFM system.

*Magneto-Optical Kerr Effect Microscopy (µMOKE) Measurements:* Kerr microscopy was used to probe the out-of-plane magnetization dynamics under applied magnetic fields. The setup operated in polar configuration, with an out-of-plane magnetic field applied by an electromagnet. P-polarized white light illuminated the sample, and the reflected beam was analyzed with an s-polarized analyzer to detect Kerr rotation. The reflected signal, collected at normal incidence, was captured using a high-resolution CCD camera. A sequence of 10 images was acquired for each value of the magnetic field to increase signal-to-noise ratio. Post-processing was performed in MATLAB to extract the hysteresis loops and the spatial maps of switching fields, exchange bias, and anisotropy. To obtain the exchange-bias anisotropy constant $K_{EB}$ from the exchange-bias field $H_{EB}$, the Stoner-Wohlfarth model for a uniformly magnetized ferromagnetic layer was used, yielding the relation: $K_{EB} = \mu_0 M_S H_{EB}$, where $M_S$ is the saturation magnetization.

*Brillouin Light Scattering (BLS):* Brillouin light scattering (BLS) measurements were performed focusing about 200 mW of a monochromatic laser beam of wavelength $\lambda = 532\ nm$, on the surface of the sample using a camera objective of numerical aperture 0.24 and focal length 50 mm. The illuminated area on the sample was a circle of approximately 40 µm in diameter, so several dozens of skyrmions were probed simultaneously. The backscattered light was frequency analyzed using a Sandercock-type (3+3)-pass tandem Fabry-Perot interferometer. BLS experiments were carried out at an angle of incidence of light $\theta = 20°$, corresponding to an in-plane wavevector $k = 4\pi \sin\theta\ ,\lambda = 8.07 \times 10^6\ rad/m$, due to the conservation of momentum in the light scattering process.

*Micromagnetic Simulations:* The micromagnetic study was performed using state-of-the-art in-house native GPU micromagnetic solver PETASPIN[57,58] which numerically integrates the Landau-Lifshitz-Gilbert-Slonczewski equation by applying the Adams-Bashforth time solver scheme:

$$\frac{d\boldsymbol{m}}{d\tau} = -\frac{1}{1+\alpha_G^2}[(\boldsymbol{m} \times \boldsymbol{h}_{\text{eff}}) - \alpha \boldsymbol{m} \times (\boldsymbol{m} \times \boldsymbol{h}_{\text{eff}})] \qquad (1)$$

where $\boldsymbol{m} = \boldsymbol{M}/M_s$ is the normalized magnetization of the ferromagnet, and $\tau = \gamma_0 M_s t$ is the dimensionless time, with $\gamma_0$ being the gyromagnetic ratio, and $M_s$ the saturation magnetization; $\alpha_G$ is the Gilbert damping. $\boldsymbol{h}_{\text{eff}}$ is the effective field normalized by the saturation field $\mu_0 M_s$, which includes the exchange, i-DMI, magnetostatic and anisotropy contributions, in addition to the external applied field. We model the Co[0.3 nm]/NiFe[1.9 nm] bilayer as a single FM film of thickness $t_{FM} = 2$ nm given the small thickness of the Co layer. The exchange bias layer is modeled via a non-uniform external field of constant magnitude (55 mT, extracted from the experimental characterization) whose orientation follows the locally imposed exchange bias direction, as shown in the left panel of Figure 4b. For instance, in the region where the ferromagnet is laser-written and the magnetization is reversed, the exchange bias field is applied with the same magnitude but opposite orientation.

We parametrize the system exploiting the experimental characterization of saturation magnetization ($M_s$), exchange stiffness, i-DMI constant ($D$) and Gilbert damping $\alpha_G$. The magnetic parameters utilized in this work are the following: $M_s$ = 720 kA/m, $A_{\text{ex}}$ = 12 pJ/m, $D$ = -0.4 mJ/m$^2$, $\alpha_G = 0.027$, and $K_u$ is varied in the range 290 - 350 kJ/m$^3$. We simulate a rectangular geometry of dimensions of 1.5 µm × 1.5 µm × $t_{FM}$ with a micromagnetic discretization in cuboid cells dimensions of 5.0 nm × 5.0 nm × $t_{FM}$; and simulations are performed at zero temperature (T = 0 K).

To excite the resonance response of the system, we apply a uniform out-of-plane time-dependent magnetic field with a *sinc* profile: $B_{\text{sinc}}(t) = B_{\text{max}} sinc(2\pi f_{\text{max}} t) =$ $B_{\max} \frac{sin(2\pi f_{max} t)}{2\pi f_{max} t}$, where $B_{\max} = 0.1$ mT is the maximum amplitude of the field and $f_{\max} = 20$ GHz is the highest excited frequency.

*Nitrogen-vacancy magnetometry*
Images of the magnetic stray fields and topography were recorded using a commercial nitrogen-vacancy (NV) scanning probe microscope (ProteusQ, Qnami AG) equipped with a diamond sensing tip hosting a single NV center located in its apex[59,60]. For topographic information, the tip is scanned in contact with the sample via frequency modulated atomic force microscopy. Simultaneously the NV Zeeman-split transitions are spectroscopically probed and tracked during scans, which yields the quantitative stray field strength along the NV axis.

**Acknowledgements**
This work was partially performed at PoliFab, the micro and nanotechnology center of the Politecnico di Milano. E.A. acknowledges funding from the European Union's Horizon 2020 research and innovation programme under grant agreement number 948225 (project B3YOND) and from the FARE programme of the Italian Ministry for University and Research (MUR) under grant agreement R20FC3PX8R (project NAMASTE). E.A. and S.T. acknowledge funding from the European Union – Next Generation EU – "PNRR – M4C2, investimento 1.1 – "Fondo PRIN 2022" – TEEPHANY– ThreEE-dimensional Processing tecHnique of mAgNetic crYstals for magnonics and nanomagnetism ID 2022P4485M CUP D53D23001400001 and CUP B53D23002820006". D.P. acknowledges funding from the European Union – Next Generation EU – "PNRR – M4C2, investimento 1.1 – "Fondo PRIN 2022" – PATH – Patterning of Antiferromagnets for THz operation id 2022ZRLA8F – CUP D53D23002490006" and from Fondazione Cariplo and Fondazione CDP, grant n° 2022-1882. AM, VP and RT acknowledge support from the projects PRIN20222N9A73 "SKYrmion-based magnetic tunnel junction to design a temperature SENSor—SkySens", and PRIN2022SAYARY "Metrology for spintronics: A machine learning approach for the reliable determination of the Dzyaloshinskii-Moriya interaction (MetroSpin)" funded by the Italian Ministry of Research, and the project PE0000021, "Network 4 Energy Sustainable Transition - NEST", funded by the European Union - Next Generation EU, under the National Recovery and Resilience Plan (NRRP), Mission 4 Component 2 Investment 1.3 - Call for Tender No. 1561 dated 11.10.2022 of the Italian MUR (CUP C93C22005230007). This work was further supported by the France 2030 government plan managed by the Agence Nationale de la Recherche (project PEPR SPIN CHIREX ANR-22-EXSP-0002 , ANR NEUROSKY ANR-23-CE24-0017-01, ACOUSKYR ANR-24-EXSP-0001).

**Laser-written reconfigurable energy landscapes and programmable Moiré spin textures**

*Matteo Panzeri, Piero Florio, Davide Girardi, Joseba Urrestarazu, Giacomo Sala, Nicola Pellizzi, Matteo Vitali, Marco Madami, Luca Ciaccarini Mavilla, Silvia Tacchi, Elisa Riedo, Andrea Meo, Riccardo Tomasello, Vito Puliafito, Efe Ersoy, Kai Wagner, Patrick Maletinsky, Olivier Boulle, Edoardo Albisetti, Daniela Petti*

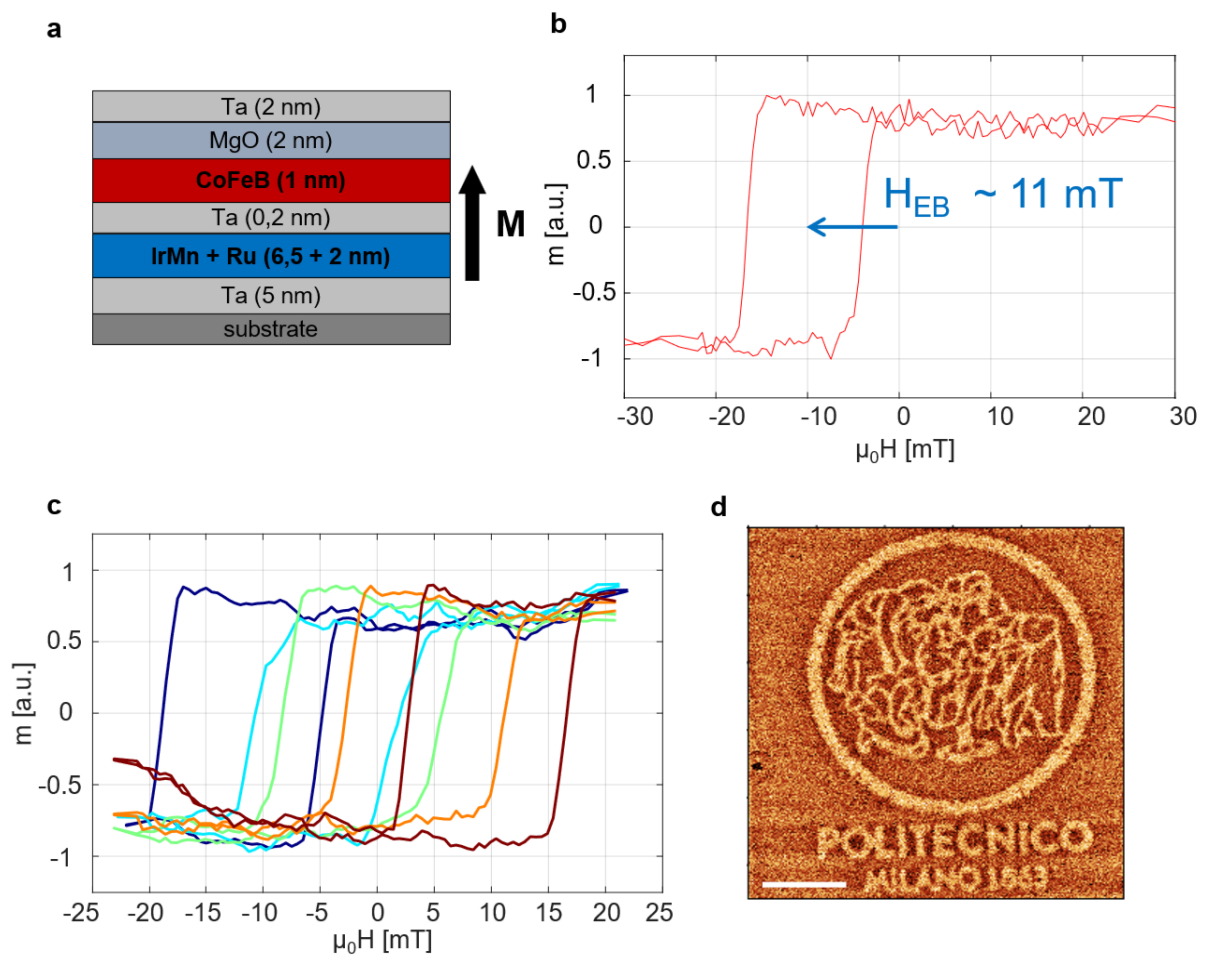

**Figure S1.** First magnetic multilayer and its characterization. a) Schematic representation of the magnetic multilayer heterostructure used in the first part of this study, deposited on a Si/SiO2 substrate. The multilayer is composed of (from bottom to top): a Ta seed layer, an IrMn antiferromagnetic (AFM) layer intercalated with Ru, a Ta interlayer, a CoFeB ferromagnetic (FM) layer, and an MgO/Ta capping layer. The structure is designed to be an exchange-biased system, where the magnetic orientation of the CoFeB layer is pinned via exchange coupling to the underlying IrMn AFM layer. The interfaces between the Ta interlayer, the CoFeB layer, and the MgO capping layer are engineered to induce a strong perpendicular magnetic anisotropy (PMA), forcing the magnetization of the CoFeB to be oriented out-of-plane (OOP). The addition of Ru to the IrMn layer serves to lower the blocking temperature of the exchange bias system. The arrow indicates the preferred direction of the magnetization. b) Representative out-of-plane (OOP) magnetic hysteresis loop of the multilayer stack, as measured by the Magneto-

Optical Kerr Effect (μMOKE) microscopy. The Kerr signal is proportional to the OOP component of the magnetization. The loop displays sharp switching characteristics and a pronounced horizontal shift towards negative magnetic fields, corresponding to an exchange bias field ($H_{EB}$) of 11 mT. Such an exchange bias value ensures that at remanence (i.e., at zero applied magnetic field), the magnetization of the ferromagnetic layer is stably pinned in a single, predetermined orientation (the "up" direction in this case), preventing magnetic reversal or domain formation. c) Local hysteresis loops corresponding to the map in Figure 1d, color-coded to match the spatial exchange bias distribution. The gradual horizontal shift of the loops along the field axis reflects the continuous modulation of the exchange bias field across the patterned region, confirming the progressive tuning of the exchange coupling induced by the laser patterning. d) MFM image of a Politecnico di Milano logo written by laser-assisted local field cooling. The pattern was produced in a single raster exposure using a constant laser power of 3.8 mW under an applied external field. The MFM contrast reproduces the intended pattern, indicating locally modified magnetic response between written and unwritten regions due to the programmed exchange bias landscape. The image demonstrates how this method allows to generate complex, high-resolution binary patterns in one pass (single-step raster writing). Scale bar: 10 μm.

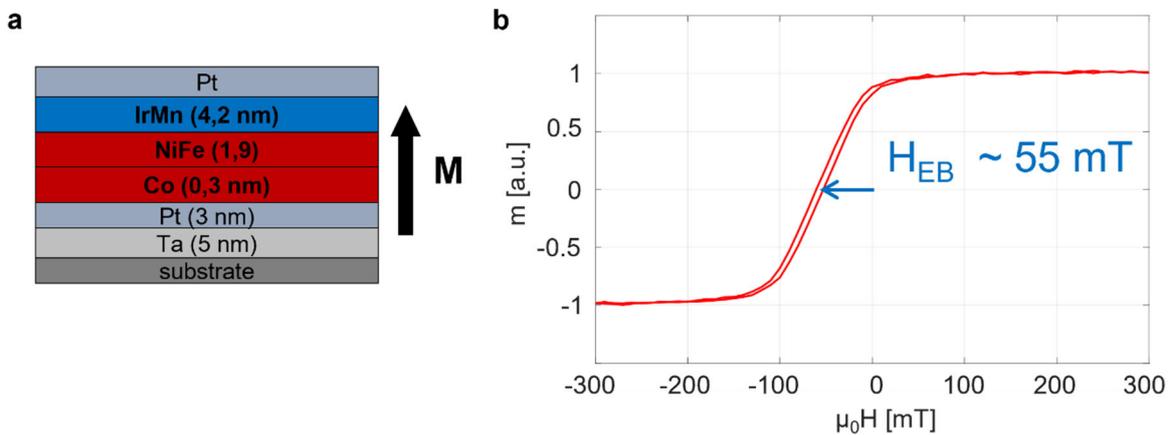

**Figure S2.** Second multilayer and its magnetic biasing. a) Schematic representation of the magnetic multilayer used in the second part of this study, deposited on a Si/SiO2 substrate. The heterostructure is composed (from bottom to top) of: Ta (seed/adhesion layer) / Pt (underlayer) / Co (ferromagnet) / NiFe (permalloy, ferromagnet) / IrMn (antiferromagnet) / Pt (capping layer). Also in this case, the IrMn provides the exchange bias, as it couples to the adjacent NiFe film. The NiFe layer is employed to obtain a large, robust exchange bias acting on the Co/Pt-based perpendicular system, characterized by a strong perpendicular magnetic anisotropy (PMA); the same interface with the heavy-metal Pt layer gives rise to an interfacial Dzyaloshinskii–Moriya interaction (DMI) and large spin-orbit torques (SOT) that are central to stabilizing and manipulating chiral spin textures. The arrow indicates the preferred direction of the magnetization. b) Out-of-plane (OOP) magnetic hysteresis loop measured by vibrating-sample magnetometry (VSM) at room temperature, plotted as normalized magnetization versus applied field. The loop exhibits a pronounced horizontal shift toward negative fields corresponding to an exchange bias field $H_{EB}$ of approximately 55 mT. Also in this sample, the exchange-bias value ensures that the ferromagnetic layer(s) are stably pinned in a single, predetermined orientation, suppressing spontaneous reversal and domain formation and thereby providing the biasing conditions used in the experiments described in the main text.

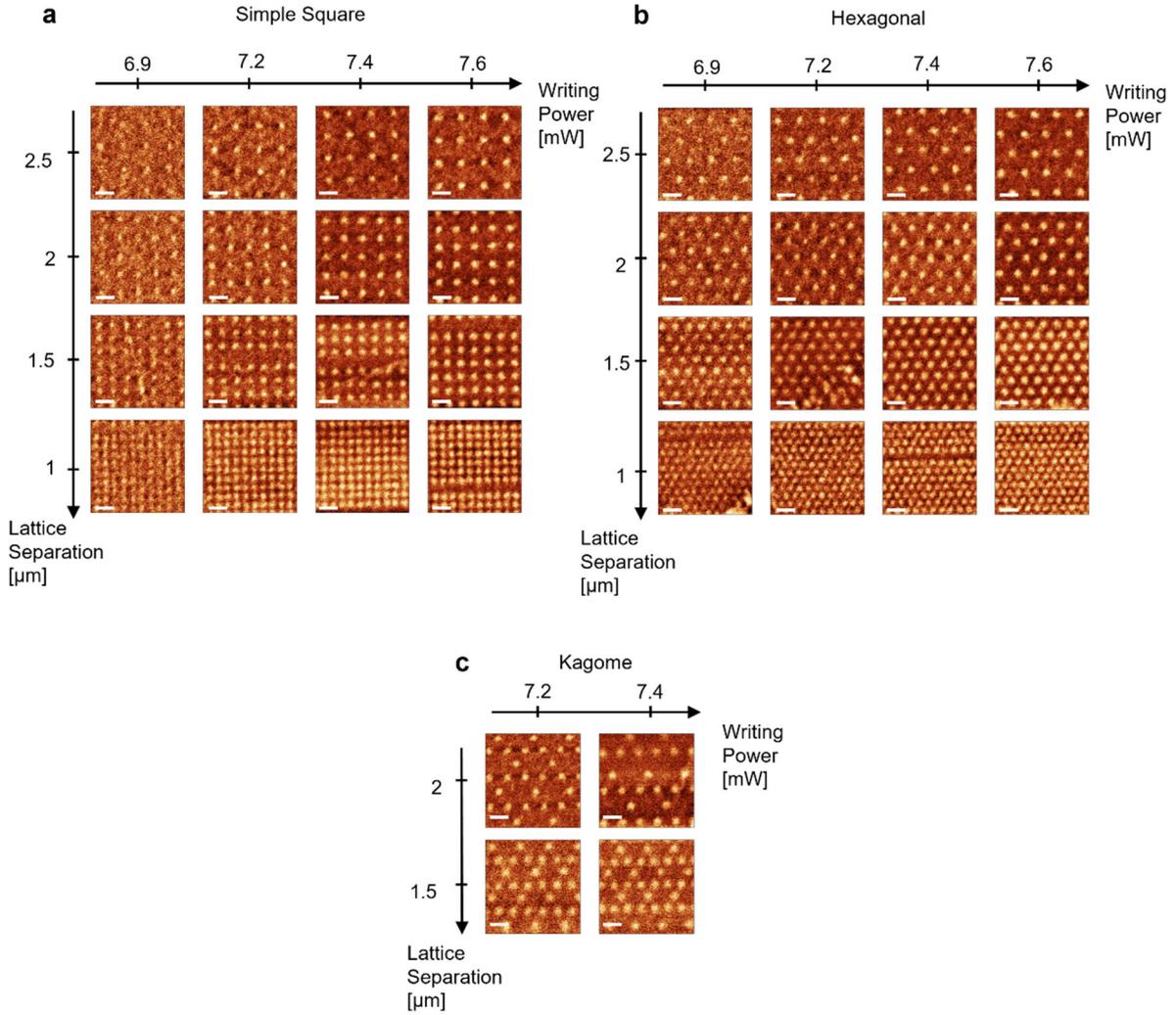

**Figure S3.** Magnetic force microscopy (MFM) images showing the deterministic writing of skyrmion lattices with programmable geometry and tunable feature size through laser-induced magnetic potential landscapes. a) Square lattices patterned with varying laser powers (6.9-7.6 mW) and lattice separations (2.5-1 µm). At the lowest power (6.9 mW), the written sites are close to the nucleation threshold and show weak MFM phase contrast, consistent with small single skyrmions (reduced stray-field gradients at the tip). Increasing the writing power enlarges the locally modified region, yielding larger skyrmions and, at the highest powers, multi-skyrmion/merged bubble domains; correspondingly, the MFM contrast increases and the lattice periodicity becomes clearer. The feature size increases from approximately 450 nm to 600 nm as the writing power is raised. b) Hexagonal skyrmion lattices patterned under equivalent conditions, demonstrating consistent control over both the lattice symmetry and skyrmion size across different geometries. c) Kagome skyrmion lattices written at 7.2 mW and 7.4 mW with lattice separations of 2 µm and 1.5 µm, respectively, further confirming the versatility of the approach in imprinting complex lattice symmetries. Across all geometries, the results highlight how the laser-programmed magnetic potential landscape defines the symmetry of the stabilized skyrmion arrays and governs the local skyrmion density per lattice site, providing deterministic control over both the size and the arrangement of the magnetic textures. All images share a scale bar of 1 µm.

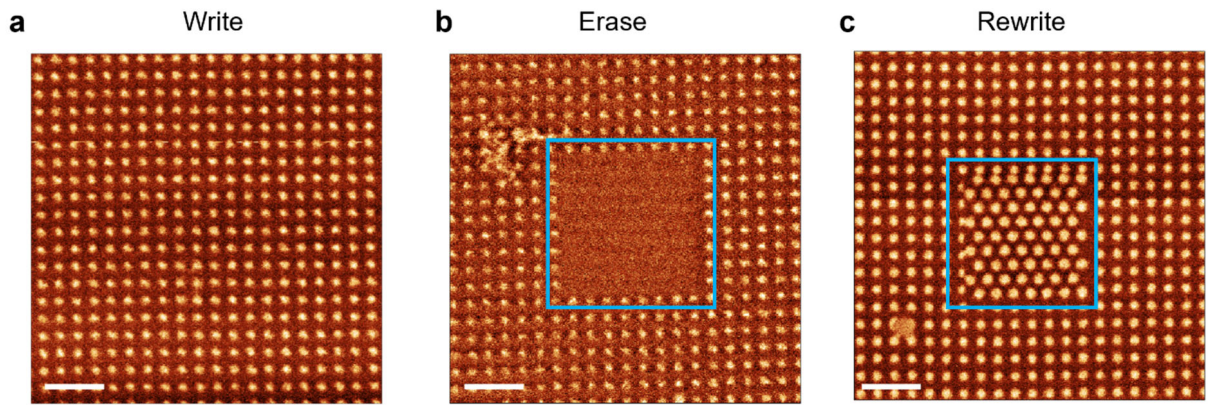

**Figure S4.** Demonstration of reversible laser writing and rewriting of skyrmion lattices. a–c) Three-step demonstration of skyrmion lattice rewritability. Panel (a) shows the MFM image of an initially written square skyrmion lattice, obtained by laser patterning under a fixed external magnetic field. Panel (b) shows the result after an erasing procedure: the first exposure is followed by a second one performed under the opposite field polarity, which selectively cancels part of the previously written pattern marked by the blue frame, confirming the reversibility of the process. Panel (c) displays the MFM image acquired after a rewriting step: after the erasing procedure, a cleared area is rewritten by switching the field direction again. In this case, a hexagonal skyrmion lattice is patterned (blue square). These results illustrate the ability of the laser-assisted programming method to locally erase and rewrite magnetic textures in a deterministic and non-destructive manner. Scale bars: 5 μm.

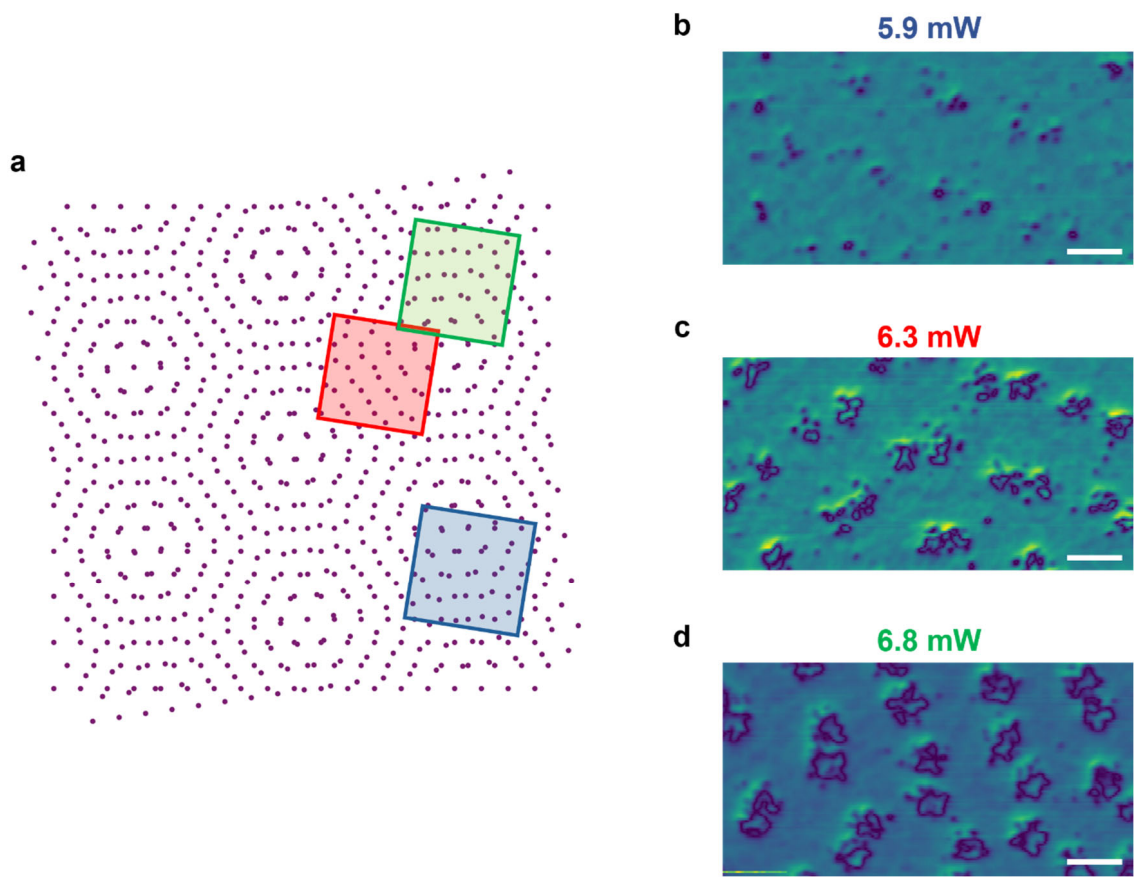

**Figure S5.** Nanoscale characterization of Moiré spin textures by NV magnetometry. a) Schematic of the laser-written Moiré lattice, with colored rectangles indicating the regions

where NV scans were performed: blue, red, and green correspond to the lattices written with laser powers of 5.9 mW, 6.3 mW, and 6.8 mW, respectively. b-d) NV magnetometry maps measured over the areas highlighted in (a), showing the evolution of the spin configuration with increasing laser power. At 5.9 mW (b), each Moiré lattice site is composed of a compact cluster of small skyrmions, rather than a single isolated one. Increasing the laser power to 6.3 mW (c) results in larger skyrmions within each cluster, while maintaining the overall Moiré periodicity. At 6.8 mW (d), the individual skyrmions further expand and partially merge, forming larger, irregular magnetic domains. These results provide direct nanoscale evidence that each Moiré lattice site corresponds to a localized ensemble of skyrmions. Unlike the sample in Figures 1-2 where laser power primarily creates a grayscale shift of the hysteresis loop, here the power tunability controls the size of the irradiated area and the local density of the stabilized skyrmions. The evolution from a dense cluster of small skyrmions at 5.9 mW (b) to larger, merged domains at 6.8 mW (d) suggests a power threshold above which a single, larger chiral domain is formed. This observation is crucial, as it indicates that the magnetic features presented in the main text (Figures 3 and 5), which were written at higher laser powers (7.4 mW and 6.8 mW, respectively), are stabilized as single chiral domains rather than skyrmion clusters. This finding validates our approach of modeling these structures in micromagnetic simulations (Figure 4) as single, localized chiral domains. Scale bars: 1 μm

**Estimation of the i-DMI via Brillouin Light Scattering (BLS)**

The strength of the i-DMI of the unpatterned multilayer has been quantitatively estimated by means of BLS measurements. In ultrathin films, the presence of i-DMI induces a frequency asymmetry between oppositely propagating Damon-Eshbach (DE) modes, which increases as a function of the wave vector k following the relation[1]:

$$\Delta f = f_{DMI}(k) - f_{DMI}(-k) = \frac{2\gamma D}{\pi M_s} k \qquad (1)$$

where $D$ is the effective DMI constant, $\gamma$ is the gyromagnetic ratio, and $M_s$ is the saturation magnetization of the magnetic layer. Using BLS, the frequency asymmetry caused by i-DMI can be quantified by measuring the frequency difference, $\Delta f$, between Stokes and anti-Stokes peaks corresponding to spin waves propagating in opposite directions. Fig.S6 shows $\Delta f$ (points) as a function of wave vector k, measured in DE geometry under an applied in-plane magnetic field $\mu_0 H_{in}$ = 300 mT and sweeping the in-plane $k$ along the perpendicular direction.
The strength of i-DMI has been determined by a linear fit (red solid line in Figure S6) of the experimental data, using equation (1). Setting the saturation magnetization to the value Ms = 720 kA/m measured by SQUID we found a D = − 0.35±0.05 mJ/m$^2$, indicating that left-handed chirality is favoured.

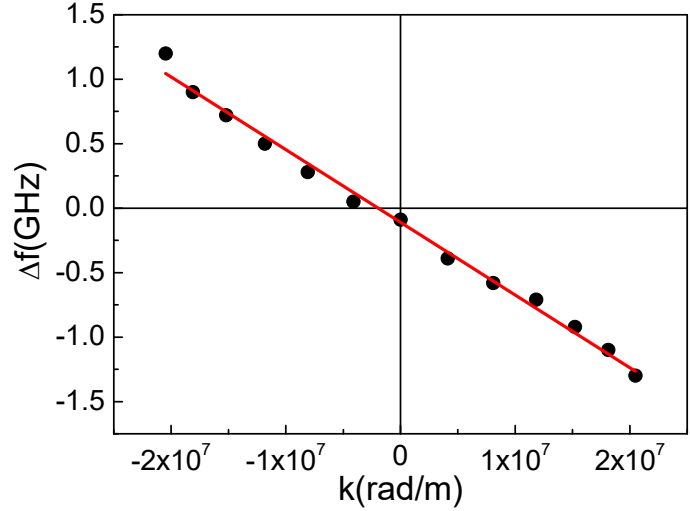

**Figure S6.** Frequency asymmetry Δ*f* (black points) between the Stokes and anti-Stokes peaks, measured as functions of k for the non-patterned multilayer. Measurements were performed in DE geometry applying an in-plane magnetic field $\mu_0 H_{in}$ = 300 mT. The red solid line represents a linear fit to the experimental data.

**Micromagnetic simulations – effect of the perpendicular magnetic anisotropy on the frequency response**

Figure S7 summarizes the effect of the perpendicular magnetic anisotropy on the system's frequency response. Panel a shows the relaxed magnetic configurations obtained for $K_u$=290, 310, 320 and 350 kJ/m$^3$, highlighting the evolution from an in-plane configuration ($K_u$=290 kJ/m$^3$) to a magnetic skyrmion in a fully out-of-plane configuration for $K_u \geq$ 320 kJ/m$^3$. This trend is confirmed by Figure S7b, which compares the field dependence of the resonant frequency for the different anisotropies ($K_u$=290, 310, 320 and 350 kJ/m$^3$) and the BLS measurements. For low anisotropy and weak fields, the system response corresponds to that of an unconfined skyrmion breathing mode, with a frequency of ~ 0.2 GHz. For larger anisotropies ($K_u \geq$ 320 kJ/m$^3$) the frequency increases to ~1.7 GHz. This is further confirmed by the frequency dependence of the normalized power spectrum of the z-component of the magnetization ($m_z$) at zero field (Figure S7c,) which shows a clear shift in the peak positions for the different anisotropies. Overall, a pronounced change in the frequency response is observed due to the change of the magnetization easy axis. For $K_u <$ 320 kJ/m$^3$, the frequency is sub-GHz at low fields, whereas for $K_u \geq$ 320 kJ/m$^3$, the frequency increases to ~1.7 GHz, in very good agreement with the experimental results. The spatially resolved two-dimensional profile of the power amplitude of $m_z$ for $K_u$ = 320 kJ/m$^3$ (Figure S7d), shows that the observed resonance corresponds to a skyrmion breathing mode, characterized by an expansion and contraction of the core/domain wall, with the mode power localized inside the circular domain wall. Inspection of the mode also reveals the effect of the confinement, which results in a not fully coherent circular excitation.

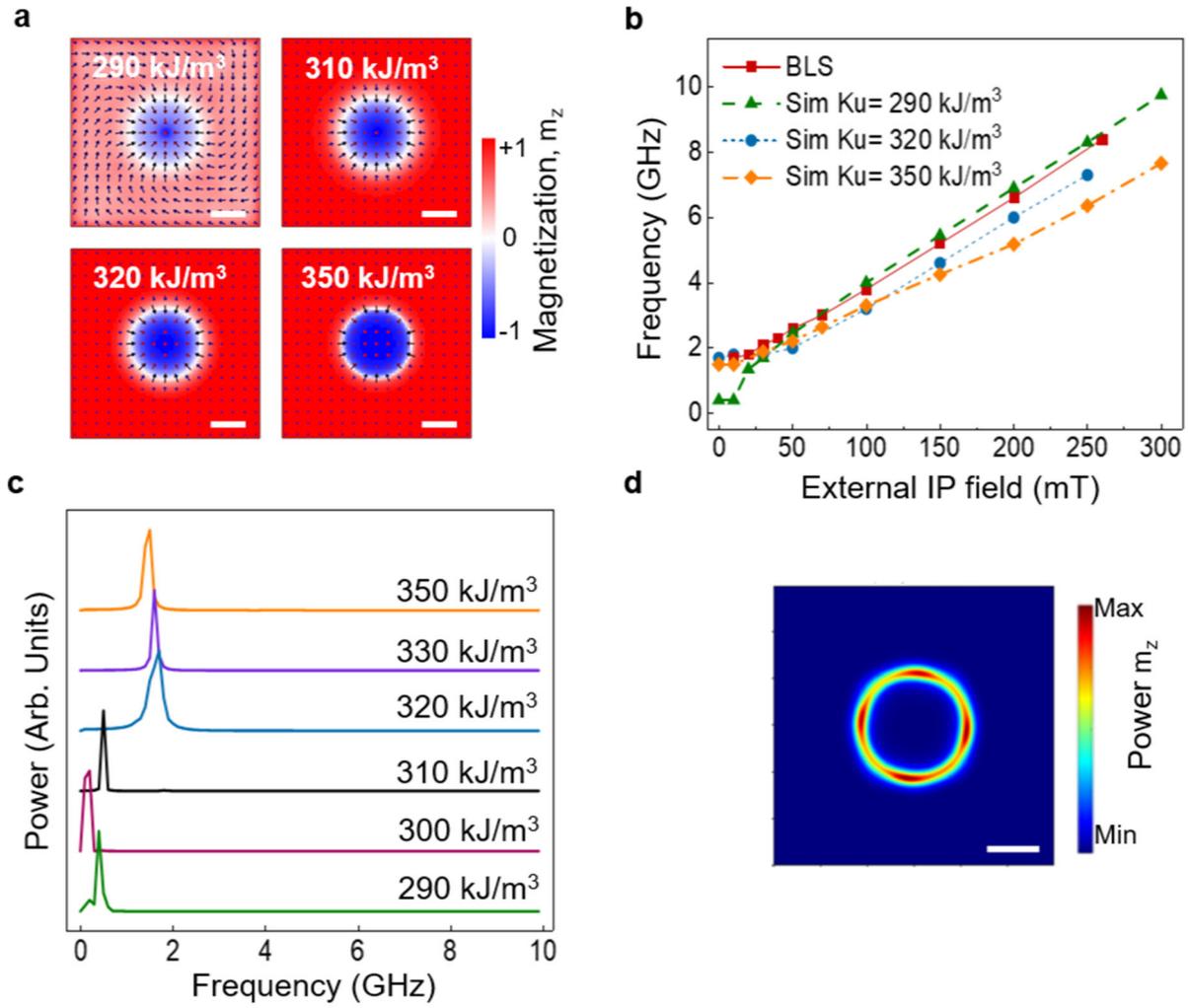

**Figure S7.** a) Snapshots of the spatial distribution of the z-component of the magnetization for the different magnetic anisotropies: $K_u$= 290, 310, 320, 350 kJ/m$^3$. The color bar represents the amplitude of the z-component of magnetization. Scale bar: 260 nm b) Average frequency of the Stokes and anti-Stokes peaks (red squares) and simulated breathing mode frequency as a function of the applied in-plane magnetic field for different $K_u$. c) Normalized power spectrum of the z-component of the magnetization as a function of frequency for different $K_u$. d) Spatial distribution of the excitation mode of the z-component of the magnetization for $K_u$ = 320 kJ/m$^3$ in zero field. The color bar gives the amplitude of the power. Scale bar: 260 nm